\documentclass[a4paper,11pt]{article}
\usepackage{amsmath}
\usepackage{amssymb}
\usepackage{cite}

\newcommand{\bm}{\boldsymbol}
\newcommand{\vect}[1]{\boldsymbol{#1}}
\newcommand{\mat}[1]{\boldsymbol{#1}}

\newcommand{\E}[1]{{\mathcal{E}}\left\{#1\right\}}
\newcommand{\ex}[1]{\exp \left\{#1\right\}}

\newcommand{\Tr}[1]{{\mathrm{Tr}}\{#1\}}
\newcommand{\Real}[1]{{\mathrm{Re}}\left[#1\right]}
\renewcommand{\det}[1]{|#1|}
\newcommand{\vectorize}[1]{\mathrm{vec}(#1)}
\newcommand{\kron}[2]{\left( #1 \otimes #2 \right)}

\newcommand{\Qt}{\mathcal{Q}_{t}}
\newcommand{\Q}{\mathcal{Q}}
\newcommand{\Rt}{\mathcal{R}_{t}}

\newcommand{\va}{\vect{a}}

\newcommand{\vu}{\vect{u}}
\newcommand{\ut}{\vect{u}_{t}}
\newcommand{\x}{\vect{x}}
\newcommand{\xt}{\vect{x}_{t}}
\newcommand{\y}{\vect{y}}
\newcommand{\z}{\vect{z}}

\newcommand{\A}{\mat{A}}
\newcommand{\B}{\mat{B}}
\newcommand{\C}{\mat{C}}
\newcommand{\F}{\mat{F}}
\newcommand{\Fg}{\mat{F}_{\text{\tiny{Gaussian}}}}
\newcommand{\Fs}{\mat{F}_{\text{\tiny{Student}}}}

\newcommand{\Hj}{\mat{H}_{j}}
\newcommand{\Hk}{\mat{H}_{k}}
\newcommand{\I}{\mat{I}}
\newcommand{\M}{\mat{M}}

\newcommand{\X}{\mat{X}}

\newcommand{\valpha}{\bm{\alpha}}
\newcommand{\vbeta}{\bm{\beta}}
\newcommand{\vtheta}{\bm{\theta}}
\newcommand{\vmu}{\bm{\mu}}
\newcommand{\mut}{\vmu_{t}}

\newcommand{\mSigma}{\bm{\Sigma}}
\newcommand{\iSigma}{\mSigma^{-1}}
\newcommand{\Sigmaj}{\mSigma_{j}}
\newcommand{\Sigmak}{\mSigma_{k}}

\newcommand{\CN}[2]{\mathbb{C}\mathcal{N}\left(#1,#2\right)}

\newcommand{\Cchisquare}[1]{\mathbb{C}\chi^{2}_{#1}}
\newcommand{\MEC}[4]{\mathbb{C}\mathcal{EC}_{#1}\left(#2,#3,#4\right)}

\newcommand{\EMS}[4]{\mathbb{C}\mathcal{EMS}_{#1}\left(#2,#3,#4\right)}
\newcommand{\EVS}[4]{\mathbb{C}\mathcal{EVS}_{#1}\left(#2,#3,#4\right)}

\newcommand{\pdfU}[1]{\mathcal{U}\left({#1}\right)}

\newcommand{\dmutdjH}{\frac{\partial \mut^{H}}{\partial \theta_j}}
\newcommand{\dmutdkH}{\frac{\partial \mut^{H}}{\partial \theta_k}}
\newcommand{\dmudjH}{\frac{\partial \vmu^{H}}{\partial \theta_j}}
\newcommand{\dmudkH}{\frac{\partial \vmu^{H}}{\partial \theta_k}}
\newcommand{\dmutdj}{\frac{\partial \mut}{\partial \theta_j}}
\newcommand{\dmutdk}{\frac{\partial \mut}{\partial \theta_k}}

\newcommand{\dmudk}{\frac{\partial \vmu}{\partial \theta_k}}
\newcommand{\dmutdjHxt}{\dmutdjH \iSigma (\xt-\mut)}
\newcommand{\dmutdkHxt}{\dmutdkH \iSigma (\xt-\mut)}
\newcommand{\qft}{(\xt-\mut)^{H} \iSigma (\xt-\mut)}
\newcommand{\qfj}{(\xt-\mut)^{H} \iSigma \Sigmaj \iSigma (\xt-\mut)}
\newcommand{\qfk}{(\xt-\mut)^{H} \iSigma \Sigmak \iSigma (\xt-\mut)}
\newcommand{\thetaj}{\theta_{j}}
\newcommand{\thetak}{\theta_{k}}

\begin{document}
\title{On the Fisher information matrix for multivariate elliptically contoured distributions}
\author{Olivier Besson\thanks{O. Besson is with the University of Toulouse, ISAE, Department Electronics Optronics Signal, 10 Avenue Edouard Belin, 31055 Toulouse, France (e-mail: olivier.besson@isae.fr)} and Yuri I. Abramovich\thanks{Y. Abramovich is with W R Systems, Ltd., 11351 Random Hills Road, Suite 400, Fairfax, VA 22030. (e-mail: yabramovich@wrsystems.com).} }
\maketitle

\begin{abstract}
The Slepian-Bangs formula provides a very convenient way to compute the Fisher information matrix (FIM) for Gaussian distributed data. The aim of this letter is to extend it to a larger family of distributions, namely elliptically contoured (EC) distributions. More precisely, we derive a closed-form expression of the FIM in this case. This new expression involves the usual term of the Gaussian FIM plus some corrective factors that depend only on the expectations of some functions of the so-called modular variate. Hence, for most distributions in the EC family, derivation of the FIM from its Gaussian counterpart involves slight additional derivations. We show that the new formula reduces to the Slepian-Bangs formula in the Gaussian case and we provide an illustrative example with Student distributions on how it can be used.
\end{abstract}

\section{Introduction}
The Cram\'{e}r-Rao bound (CRB) provides a lower bound on the variance of any unbiased estimator and is thus the ubiquitous reference to compare the performance of a given estimator to \cite{Kay93}. The CRB is usually computed as the inverse of the Fisher information matrix (FIM) whose entries involve the derivatives of the log-likelihood function of the observation matrix $\X=\begin{bmatrix} \x_{1} & \x_{2} & \ldots & \x_{T} \end{bmatrix}$ where $\xt \in \mathbb{C}^{M}$ stands for the $t$-th snapshot. When the latter are independent and identically distributed (i.i.d.) vectors drawn from a complex Gaussian distribution, i.e., when $\xt \sim \CN{\mut(\vtheta)}{\mSigma (\vtheta)}$ where $\vtheta \in \mathbb{R}^{p}$ denotes the set of unknown real-valued parameters that describe the distribution, the Slepian-Bangs formula \cite{Slepian54,Bangs71} provides a general expression for the FIM as
\begin{align}\label{Slepian-Bangs}
\F(j,k) &= \E{\frac{\partial \log p(\X | \vtheta)}{\partial \thetaj} \frac{\partial \log p(\X | \vtheta)}{\partial \thetak}} = \E{-\frac{\partial^{2} \log p(\X | \vtheta)}{\partial \thetaj \partial \thetak}} \nonumber \\
&= 2 \sum_{t=1}^{T} \Real{\dmutdjH \iSigma \dmutdkH} + T \Tr{\iSigma \Sigmaj \iSigma \Sigmak}
\end{align}
where $\Sigmaj = \dfrac{\partial \mSigma}{\partial \thetaj}$ and where, for the sake of simplicity, we have omitted the dependence of $\mut$ and $\mSigma$ on $\vtheta$ in the second line of \eqref{Slepian-Bangs}. The convenience of such formula has been thoroughly used for a myriad of statistical data models, at least under the Gaussian framework. However, in many applications, non-Gaussianity of the data has been evidenced and hence, for each non-Gaussian distribution, the FIM must be specifically computed from the first line of \eqref{Slepian-Bangs}. In this letter, we provide an extension of the Slepian-Bangs formula to a very general class of distributions, namely multivariate elliptically contoured (EC) distributions. As will be shown shortly, a very simple formula, similar to \eqref{Slepian-Bangs}, can be obtained. The new formula involves the two terms of the Slepian-Bangs formula with some scaling factors, and the latter can be expressed simply as the statistical expectation of some functions of the so-called modular variate. These expectations are usually simple to obtain (see illustrative examples below) and hence it becomes rather straightforward to derive the FIM for EC distributions from the FIM for Gaussian distributions.

\section{A brief review of EC distributions}
Multivariate elliptically contoured distributions \cite{Anderson82,Fang84,Krishnaiah86,Fang90,Anderson90,Micheas06} constitute a large family of distributions which have been used in a variety of applications, including array processing. In this section, we  briefly summarize their definitions and properties so as to provide the necessary background for derivation of the FIM in the next sections. A very detailed presentation of EC distributions can be found in the book \cite{Fang90} which constitutes the most cited reference on this topic. We would like also to point to the recent paper \cite{Ollila12b} where a very comprehensive review of complex elliptically symmetric distributions is given, along with results on estimation within this framework. The reader is referred to these references for some  details that could be skipped in the short review to be presented now. A vector $\xt \in \mathbb{C}^{M}$  follows an EC distribution if it admits the following stochastic representation
\begin{equation}\label{storep_CES}
\xt \overset{d}{=} \mut + \Rt \C \ut
\end{equation}
where the non-negative real random variable $\Rt = \sqrt{\Qt}$, called the modular variate,  is independent of the complex random vector $\ut$ possessing  a uniform distribution on the complex sphere $\mathbb{C}S^{R} = \left\{ \z \in \mathbb{C}^{R}; \left\| \z \right\|=1 \right\}$, which we denote as $\ut \sim \pdfU{\mathbb{C}S^{R}}$. In \eqref{storep_CES}, $\overset{d}{=}$ means ``has the same distribution as''. The full-rank matrix $\C \in \mathbb{C}^{M \times R}$ is such that $\C \C^{H} = \mSigma$ where $\mSigma$ is the so-called scatter matrix. In this paper, we consider the special absolutely continuous case where $\mSigma$ is non singular and hence $R=M$. In such a case, the probability density function (p.d.f.) of $\xt$ can be defined and is given by
\begin{equation}\label{p(xt)}
p(\xt | \mut, \mSigma, g) = C_{M,g} \det{\mSigma_{0}}^{-1} g \left( \qft \right)
\end{equation}
for some function $g: \, \mathbb{R}^{+} \, \longrightarrow \, \mathbb{R}^{+}$ called density generator that satisfies finite moment condition $\delta_{M,g} = \int_{0}^{\infty} t^{M-1} g(t) dt < \infty$. The density generator is related to the p.d.f. of the modular variate $\Qt$ by 
\begin{equation}\label{p(Qt)}
p(\Qt) = \delta_{M,g}^{-1} \Qt^{M-1} g(\Qt).
\end{equation}
We adopt the following notation in the following $\xt \sim \MEC{M}{\mut}{\mSigma}{g}$. The complex Gaussian distribution $\CN{\vect{\mut}}{\mSigma}$ is obtained for the particular $g(t)=\ex{-t}$. 

While there is essentially a unique way to define an elliptically contoured distribution for a vector, when it comes to extend it to the matrix-variate   $\X =\begin{bmatrix} \x_{1} & \x_{2} & \cdots & \x_{T} \end{bmatrix} \in \mathbb{C}^{M \times T}$, several options are possible \cite{Fang90}. Indeed, Fang and Zhang distinguish four classes of matrix-variate elliptically contoured distributions whose p.d.f. and stochastic representations are different. In this paper, we will focus on the two main matrix-variate distributions encountered in the array processing literature, namely (in the terminology of \cite{Fang90,Anderson90})
\begin{enumerate}
\item the multivariate elliptical distributions \cite{Anderson82} where essentially all snapshots $\xt$ are independent and identically distributed  according to \eqref{p(xt)}. We will denote this type of distribution as $ \X \sim \EMS{M \times T}{\M}{\mSigma}{g}$ where $\M = \begin{bmatrix} \vmu_{1} & \vmu_{2} & \ldots & \vmu_{T} \end{bmatrix}$.
\item the vector elliptical distributions where $\x = \vectorize{\X} = \begin{bmatrix} \x_{1}^{T} & \cdots & \x_{T}^{T} \end{bmatrix}^{T} \in \mathbb{C}^{M T}$ follows a (vector) EC distribution, i.e., $\x \overset{d}{=} \vmu + \sqrt{\Q} \kron{\I_{T}}{\mSigma^{1/2}} \vu$ with $\vmu = \begin{bmatrix} \vmu_{1}^{T} & \vmu_{2}^{T} & \ldots & \vmu_{T}^{T} \end{bmatrix}^{T}$ and $\vu \sim \pdfU{\mathbb{C}S^{MT}}$. We denote this distribution as $ \X \sim \EVS{M \times T}{\M}{\mSigma}{g}$.
\end{enumerate}

\section{The FIM for EMS distributions}
Let us investigate first the EMS type of distributions. The latter have been considered for instance in radar applications in order to model clutter. Indeed, in many radar scenarios,  clutter has been evidenced to be non-Gaussian and hence a lot of studies have focused on clutter distribution modeling and assessment. One of the most popular models so far is the so-called compound-Gaussian model where the clutter returns are modeled as spherically invariant random vectors \cite{Yao73,Conte87,Conte95}. The latter belong to the larger class of EC distributed data. Within this framework, a great deal of attention has focused on estimation of the scatter matrix, see e.g., \cite{Gini02,Abramovich07c,Pascal08,Chen11,Wiesel12}. Note that Kent and Tyler in the eighties addressed a similar framework in the statistical literature \cite{Tyler87,Tyler87b,Kent88,Kent91}. As we said before, we assume that  the $T$ snapshots $\xt$ are  i.i.d random vectors  drawn from $\xt \sim \MEC{M}{\mut}{\mSigma}{g}$. Therefore, the p.d.f. of $\X =\begin{bmatrix} \x_{1} & \x_{2} & \cdots & \x_{T} \end{bmatrix} \in \mathbb{C}^{M \times T}$ is given by
\begin{equation}\label{p(X)_EMS}
p(\X | \vmu_{1},\ldots,\vmu_{T}, \mSigma, g ) = C_{M,g}^{T} \det{\mSigma}^{-T} \prod_{t=1}^{T} g(\qft).
\end{equation}
In the sequel, we assume that $\vmu_{1},\ldots,\vmu_{T}$ and $\mSigma$ depend on an unknown parameter vector $\vtheta$, which we wish to estimate from $\X$, and we look for an expression for the FIM under this statistical model. For the sake of convenience, we rewrite the likelihood function in \eqref{p(X)_EMS} as $p(\X | \vtheta , g)$ and we will omit the explicit dependence of $\mut$ and $\mSigma$ on $\vtheta$.

In order to obtain the FIM, we must first compute the first-order derivative of the log-likelihood function
\begin{equation}\label{L(X)_EMS}
L(\X | \vtheta, g ) = \mathrm{const}. - T \log \det{\mSigma} + \sum_{t=1}^{T} \log g(\eta_t)
\end{equation}
where $\eta_t \overset{\text{def}}{=} \qft$. Differentiating \eqref{L(X)_EMS} with respect to (w.r.t.) $\thetaj$, we obtain
\begin{equation}\label{dLdthetaj_EMS}
\frac{\partial L(\X | \vtheta, g )}{\partial \thetaj} = - T \Tr{\iSigma \Sigmaj} + 
\sum_{t=1}^{T} \phi(\eta_t) \frac{\partial \eta_t}{\partial \thetaj}
\end{equation}
where $\phi(t) \overset{\text{def}}{=} \dfrac{g'(t)}{g(t)}$. Now,
\begin{align}\label{detatdthetaj}
\frac{\partial \eta_t}{\partial \thetaj} &= -2 \Real{\dmutdjHxt} \nonumber \\
&- \qfj.
\end{align}
Let us first prove that
\begin{equation}\label{cond_LL_EMS}
\E{\frac{\partial L(\X | \vtheta, g )}{\partial \thetaj}} = 0
\end{equation}
which is a necessary condition for the CRB theory to apply. Making use of $\xt \overset{d}{=} \mut + \sqrt{\Qt} \mSigma^{1/2} \ut$, one can observe that $\eta_t \overset{d}{=} \Qt$  and $\qfj \overset{d}{=} \Qt \left(\ut^{H} \Hj \ut\right)$ with $\Hj \overset{\text{def}}{=} \mSigma^{-1/2} \Sigmaj \mSigma^{-1/2}$. Therefore, 
\begin{align}
\E{\phi(\eta_t) \frac{\partial \eta_t}{\partial \thetaj}} &= - 2 \E{\Qt^{1/2} \phi(\Qt)  \Real{\dmutdjH \mSigma^{-1/2} \ut}} \nonumber \\
&- \E{\Qt \phi(\Qt) \left[ \ut^{H} \Hj \ut \right]}. 
\end{align}
Now, $\Qt$ and $\ut$ are independent. Moreover, $\ut \sim \pdfU{\mathbb{C}S^{M}}$ and hence $\E{\ut} = \vect{0}$ and $\E{\ut \ut^{H}} = M^{-1} \I_{M}$. Furthermore
\begin{align}
\E{\Qt \phi(\Qt)} &= \int_{0}^{\infty} \delta_{M,g}^{-1} \Qt^{M} g'(\Qt) d \Qt \nonumber \\
&= \left[ \delta_{M,g}^{-1} \Qt^{M} g(\Qt) \right]^{\infty}_{0} -  M \int_{0}^{\infty} \delta_{M,g}^{-1} \Qt^{M-1} g(\Qt) d \Qt \nonumber \\
&= -M
\end{align}
and
\begin{equation}
\E{\ut^{H} \Hj \ut} = \E{\Tr{\ut \ut^{H} \Hj}} = M^{-1} \Tr{\Hj}.
\end{equation}
It ensues that
\begin{equation}
\E{\phi(\eta_t) \frac{\partial \eta_t}{\partial \thetaj}} = \Tr{\Hj} = \Tr{\iSigma \Sigmaj}.
\end{equation}
Reporting this equation in \eqref{dLdthetaj_EMS} proves \eqref{cond_LL_EMS}.

Let us now turn to the derivation of the $(j,k)$ entry of the FIM:
\begin{align}
\F(j,k) &= \E{\frac{\partial \log p(\X | \vtheta)}{\partial \thetaj} \frac{\partial \log p(\X | \vtheta)}{\partial \thetak}} \nonumber \\
&= -T^{2} \Tr{\iSigma \Sigmaj} \Tr{\iSigma \Sigmak} + \E{\sum_{t,s=1}^{T} \phi(\eta_t) \phi(\eta_s) \frac{\partial \eta_t}{\partial \thetaj} \frac{\partial \eta_s}{\partial \thetak} } \nonumber \\
&= -T^{2} \Tr{\iSigma \Sigmaj} \Tr{\iSigma \Sigmak} + \sum_{t \neq s} \Tr{\iSigma \Sigmaj} \Tr{\iSigma \Sigmak} \nonumber \\
&+ \E{\sum_{t=1}^{T} \phi^{2}(\eta_t) \frac{\partial \eta_t}{\partial \thetaj} \frac{\partial \eta_t}{\partial \thetak}} \nonumber \\
&= -T \Tr{\iSigma \Sigmaj} \Tr{\iSigma \Sigmak} + \E{\sum_{t=1}^{T} \phi^{2}(\eta_t) \frac{\partial \eta_t}{\partial \thetaj} \frac{\partial \eta_t}{\partial \thetak}}.
\end{align}
Now, we have
\begin{align}\label{detadjxdetadk}
\frac{\partial \eta_t}{\partial \thetaj} \frac{\partial \eta_t}{\partial \thetak} &= 4 \Real{\dmutdjHxt} \Real{\dmutdkHxt} \nonumber \\
&+ 2 \Real{\dmutdjHxt} \left[\qfk\right] \nonumber \\
& + 2 \Real{\dmutdkHxt} \left[\qfj\right] \nonumber \\
&+ \left[\qfj\right] \left[\qfk\right].
\end{align}
Therefore, we need to evaluate the expected value of the three different terms in the previous equation, which we do now. More precisely, using $\mSigma^{1/2} (\xt - \mut) \overset{d}{=} \sqrt{\Qt} \ut$, one has
\begin{align}\label{T1}
T_{1} &= \E{\phi^{2}(\eta_t) \Real{\dmutdjHxt} \Real{\dmutdkHxt}} \nonumber \\
&= \E{\Qt \phi^{2}(\Qt)} \E{\Real{\dmutdjH \mSigma^{-1/2} \ut} \Real{\dmutdkH \mSigma^{-1/2} \ut}} \nonumber \\
&= \frac{1}{4} \E{\Qt \phi^{2}(\Qt)} \E{\left[ \dmutdjH \mSigma^{-1/2} \ut + \ut^{H} \mSigma^{-1/2} \dmutdj  \right] \left[ \dmutdkH \mSigma^{-1/2} \ut + \ut^{H} \mSigma^{-1/2} \dmutdk  \right]} \nonumber \\
&= \frac{1}{4} \E{\Qt \phi^{2}(\Qt)}   \E{\dmutdjH \mSigma^{-1/2} \ut \ut^{H} \mSigma^{-1/2} \dmutdk + \dmutdkH \mSigma^{-1/2} \ut \ut^{H} \mSigma^{-1/2} \dmutdj } \nonumber \\
&= \frac{1}{4M} \E{\Qt \phi^{2}(\Qt)} \left[ \dmutdjH \iSigma \dmutdk + \dmutdkH \iSigma \dmutdj \right] \nonumber \\
&= \frac{1}{2M} \E{\Qt \phi^{2}(\Qt)} \Real{\dmutdjH \iSigma \dmutdk}.
\end{align}
Let us next address the second term:
\begin{align}
T_{2} &= \E{\phi^{2}(\eta_t) \Real{\dmutdjHxt} \left[\qfk\right]} \nonumber \\
&= \frac{1}{2}\E{\Qt^{3/2} \phi^{2}(\Qt)} 
\times \E{\left[ \dmutdjH \mSigma^{-1/2} \ut + \ut^{H} \mSigma^{-1/2} \dmutdj  \right] \left[\ut^{H} \Hk \ut \right]}.
\end{align}
At this stage, we need to compute $\E{(\ut^{H} \va) (\ut^{H} \B \ut)}$ for some vector $\va$ and Hermitian matrix $\B$. Towards this end, let $\y \sim \CN{\vect{0}}{\I_{M}}$ and note that $\y = \left\| \y \right\| \vu$ where $\left\| \y \right\|^{2} \sim \Cchisquare{M}$ and $\vu \sim \pdfU{\mathbb{C}S^{M}}$ are independent. Since $\y$ is Gaussian distributed, one has $\E{(\y^{H} \va) (\y^{H} \B \y)}=0$. However, 
$\E{(\y^{H} \va) (\y^{H} \B \y)} = \E{\left\| \y \right\|^{3}} \E{(\ut^{H} \va) (\ut^{H} \B \ut)}$ and $\E{\left\| \y \right\|^{3}} > 0$. Therefore, $\E{(\ut^{H} \va) (\ut^{H} \B \ut)}=0$ and hence $T_{2}=0$.

It remains to derive the last term in \eqref{detadjxdetadk}, namely
\begin{align}
T_{3} &= \E{\left[\qfj\right] \left[\qfk\right]} \nonumber \\
&= \E{\Qt^{2} \phi^{2}(\Qt)} \E{\left[\ut^{H} \Hj \ut\right] \left[\ut^{H} \Hk \ut\right]}.
\end{align}
Similarly to what was done before, let us consider $\y \sim \CN{\vect{0}}{\I_{M}}$ with $\y = \left\| \y \right\| \vu$. It is well known that
\begin{subequations}
\begin{align}
\E{\left[\y^{H} \A \y\right] \left[\y^{H} \B \y \right]} &= \Tr{\A} \Tr{\B} + \Tr{\A \B} \\
\E{\left\| \y \right\|^{4}} &= \E{\left( \Cchisquare{M} \right)^{2}} = M(M+1).
\end{align}
\end{subequations}
Consequently
\begin{equation}
\E{\left[\vu^{H} \A \vu\right] \left[\vu^{H} \B \vu\right]} = \frac{\Tr{\A}\Tr{\B} + \Tr{\A \B}}{M(M+1)}
\end{equation}
from which we infer that
\begin{equation}\label{T3}
T_{3} = \frac{\E{\Qt^{2} \phi^{2}(\Qt)}}{M(M+1)} \left[ \Tr{\iSigma \Sigmaj} \Tr{\iSigma \Sigmak} + \Tr{\iSigma \Sigmaj \iSigma \Sigmak} \right].
\end{equation}
Using \eqref{detadjxdetadk}, \eqref{T1} and \eqref{T3} in the expression of the FIM, we finally obtain the following \textbf{extension of the Slepian-Bangs formula to EMS distributions}:
\begin{align}\label{FIM_EMS}
\F(j,k) &= -T \Tr{\iSigma \Sigmaj} \Tr{\iSigma \Sigmak} \nonumber  \\
&+ \frac{2}{M} \sum_{t=1}^{T} \E{\Qt \phi^{2}(\Qt)} \Real{\dmutdjH \iSigma \dmutdk} \nonumber \\
& + \frac{\Tr{\iSigma \Sigmaj} \Tr{\iSigma \Sigmak} + \Tr{\iSigma \Sigmaj \iSigma \Sigmak}}{M(M+1)} \sum_{t=1}^{T} \E{\Qt^{2} \phi^{2}(\Qt)}.
\end{align}

It is remarkable that despite the high generality of EC distributions, the formula for the FIM remains quite simple. Indeed, it is reminiscent of the FIM for Gaussian distributions (one recognizes the two terms of the Slepian-Bangs formula) but for different scaling factors. The latter depend only on the expected values of some functions of the modular variate, and deviation from the Gaussian distribution manifests itself only through these terms. Note that the latter involve only scalar integrals and hence, in many cases, one might expect an analytic expression for them. Would that not be the case, currently available numerical tools enable one to compute the required integrals. This means that any Fisher information matrix derived under the Gaussian assumption needs to be modified only slightly to obtain the FIM for EMS distributions: indeed, only computation of $\E{\Qt \phi^{2}(\Qt)}$ and $\E{\Qt^{2} \phi^{2}(\Qt)}$ is necessary. This property paves the way to extension of many FIM derived so far under the Gaussian umbrella.

We also observe that if $\mSigma$ is known, then the FIM for EMS distributions is directly proportional to the Gaussian FIM: hence, non-Gaussianity results in scaling of the CRB. Accordingly, if $\vtheta = \begin{bmatrix} \valpha^{T} & \vbeta^{T} \end{bmatrix}^{T}$ where $\mut$ depends only on $\valpha$ and $\mSigma$ depends only on $\vbeta$, then the FIM is block-diagonal. Moreover, the FIM for estimation of $\valpha$ only is proportional to the Gaussian FIM, a fact that was already discovered in \cite{Sadler99}.  In contrast, when $\mut = \vect{0}$ the two FIM are no longer proportional, due to the term $\Tr{\iSigma \Sigmaj} \Tr{\iSigma \Sigmak}$.

We now provide illustrative examples of how this formula can be used. Of course, we start with the Gaussian assumption for which $g(t)=\ex{-t}$ and 
\begin{equation*}
p(\Qt) = \frac{1}{\Gamma(M)} \Qt^{M-1} \ex{-\Qt}.
\end{equation*}
In this case, we have $\phi(t)=-1$ and
\begin{subequations}
\begin{align}
\E{\Qt \phi^{2}(\Qt)} &= M \\
\E{\Qt^{2} \phi^{2}(\Qt)} &= M(M+1).
\end{align}
\end{subequations}
Reporting this value in \eqref{FIM_EMS} yields
\begin{align}\label{FIM_Gaussian}
\Fg(j,k) &= -T \Tr{\iSigma \Sigmaj} \Tr{\iSigma \Sigmak} \nonumber  \\
&+ 2 \sum_{t=1}^{T} \Real{\dmutdjH \iSigma \dmutdk} \nonumber \\
& + T \left[\Tr{\iSigma \Sigmaj} \Tr{\iSigma \Sigmak} + \Tr{\iSigma \Sigmaj \iSigma \Sigmak} \right]
\end{align}
which coincides with the Slepian-Bangs formula \eqref{Slepian-Bangs}.

Let us now consider the well-known Student distribution with $d$ degrees of freedom given by
\begin{equation}
p_{\text{\tiny{Student}}}(\xt) = \frac{\Gamma(d+M)}{\pi^{M} d^{M} \Gamma(d)} \det{\mSigma}^{-1} \left[ 1 + d^{-1} \qft \right]^{-(d+M)}.
\end{equation}
This corresponds to $g(t)=(1+d^{-1}t)^{-(d+M)}$ and hence $\phi(t)=- \frac{d+M}{d}(1+d^{-1}t)^{-1}$. Moreover, $\Qt \overset{d}{=} \dfrac{\Cchisquare{M}}{\Cchisquare{d}/d}$, and hence $\Qt$ follows a scaled $\mathcal{F}$-distribution:
\begin{equation}
p(\Qt) = \frac{\Gamma(d+M)}{d^{M} \Gamma(d) \Gamma(M)} \Qt^{M-1} (1+d^{-1}\Qt)^{-(d+M)}.
\end{equation}
Some straightforward calculations show that, in this case
\begin{subequations}
\begin{align}
\E{\Qt \phi^{2}(\Qt)} &= \frac{(d+M)M}{d+M+1} \\
\E{\Qt^{2} \phi^{2}(\Qt)} &= \frac{(d+M)M(M+1)}{d+M+1}.
\end{align}
\end{subequations}
Consequently, in the Student case, the FIM has the following expression
\begin{align}\label{FIM_Student}
\Fs(j,k) &= -T \Tr{\iSigma \Sigmaj} \Tr{\iSigma \Sigmak} \nonumber  \\
&+ 2 \frac{d+M}{d+M+1} \sum_{t=1}^{T} \Real{\dmutdjH \iSigma \dmutdk} \nonumber \\
& + \frac{(d+M)T}{d+M+1} \left[\Tr{\iSigma \Sigmaj} \Tr{\iSigma \Sigmak} + \Tr{\iSigma \Sigmaj \iSigma \Sigmak} \right].
\end{align}
One can verify, as expected, that $\lim_{d \rightarrow \infty} \Fs = \Fg$.

\section{The FIM for EVS distributions}
Let us now consider the case where 
\begin{equation}
\x \overset{d}{=} \vmu + \sqrt{\Q} \kron{\I_{T}}{\mSigma^{1/2}} \vu
\end{equation}
with $\vu \sim \pdfU{\mathbb{C}S^{MT}}$. This model has been used in the array processing context, e.g., in \cite{Richmond96c,Richmond96d} where Christ Richmond investigated the extension of well-known detection schemes developed in the Gaussian framework (viz. Kelly's generalized likelihood ratio test \cite{Kelly86}) to EC distributions. Very interestingly, Richmond proved the nice result that Kelly's detector remains the generalized likelihood ratio test for the EVS type of distribution. We now have the p.d.f. of $\X$ as
\begin{align}\label{p(X)_EVS}
p(\X | \vtheta, g ) &= C_{MT,g} \det{\mSigma}^{-T} g \left( \sum_{t=1}^{T}\qft \right) \nonumber \\
&=  C_{MT,g} \det{\mSigma}^{-T} g \left( (\x-\vmu)^{H} \kron{\I_{T}}{\iSigma} (\x-\vmu) \right).
\end{align}
Similarly to the previous section, we let $\eta_t = \qft $ and $\eta = \sum_{t=1}^{T} \eta_t \overset{d}{=} \Q$. The log-likelihood function is now
\begin{equation}\label{L(X)_EVS}
L(\X | \vtheta, g ) = \mathrm{const}. - T \log \det{\mSigma} + \log g \left(  \sum_{t=1}^{T} \eta_t  \right).
\end{equation}
Differentiating \eqref{L(X)_EVS} with respect to  $\thetaj$ yields
\begin{align}\label{dLdthetaj_EVS}
\frac{\partial L(\X | \vtheta, g )}{\partial \thetaj} &= - T \Tr{\iSigma \Sigmaj} + \phi(\eta) \sum_{t=1}^{T}  \frac{\partial \eta_t}{\partial \thetaj} \nonumber \\
&= - T \Tr{\iSigma \Sigmaj} -2 \phi(\eta)  \Real{\dmudjH \kron{\I_{T}}{\iSigma} (\x-\vmu)} \nonumber \\
& - \phi(\eta) \left[ (\x-\vmu)^{H} \kron{\I_{T}}{\iSigma \Sigmaj \iSigma} (\x-\vmu) \right]
\end{align}
where we have used \eqref{detatdthetaj}. Let us again prove that
\begin{equation}\label{cond_LL_EVS}
\E{\frac{\partial L(\X | \vtheta, g )}{\partial \thetaj}} = 0.
\end{equation}
Since $\kron{\I_{T}}{\mSigma^{-1/2}}(\x-\vmu) \overset{d}{=} \sqrt{Q} \vu$, it follows that
\begin{align}
\E{\phi(\eta)  \sum_{t=1}^{T}  \frac{\partial \eta_t}{\partial \thetaj}} &= - 2 \E{\Q^{1/2} \phi(\Q) \dmudjH \kron{\I_{T}}{\mSigma^{-1/2}} \vu} \nonumber \\
&- \E{\Q \phi(\Q) \left[ \vu^{H} \kron{\I_{T}}{\Hj} \vu \right]} \nonumber \\
&=- \E{\Q \phi(\Q)} \E{ \vu^{H} \kron{\I_{T}}{\Hj} \vu } \nonumber \\
&= (MT) \times (MT)^{-1} \Tr{\kron{\I_{T}}{\Hj}} \nonumber \\
&= T \Tr{\iSigma \Sigmaj}
\end{align}
which, when reported in \eqref{dLdthetaj_EVS}  proves \eqref{cond_LL_EVS}. The $(j,k)$ entry of the FIM can thus be written as
\begin{align}
\F(j,k) &= \E{\frac{\partial \log p(\X | \vtheta)}{\partial \thetaj} \frac{\partial \log p(\X | \vtheta)}{\partial \thetak}} \nonumber \\
&= -T^{2} \Tr{\iSigma \Sigmaj} \Tr{\iSigma \Sigmak} + \E{\phi^{2}(\eta) \left( \sum_{t=1}^{T}  \frac{\partial \eta_t}{\partial \thetaj}\right) \left( \sum_{s=1}^{T}  \frac{\partial \eta_s}{\partial \thetak}\right) }.
\end{align}
Now, we have
\begin{align}
 \left( \sum_{t=1}^{T}  \frac{\partial \eta_t}{\partial \thetaj}\right) \left( \sum_{s=1}^{T}  \frac{\partial \eta_s}{\partial \thetak}\right) &\overset{d}{=} 4 \Q \Real{\dmudjH \kron{\I_{T}}{\mSigma^{-1/2}} \vu}  \Real{\dmudkH \kron{\I_{T}}{\mSigma^{-1/2}} \vu}  \nonumber \\
 &+ 2 \Q^{3/2} \Real{\dmudjH \kron{\I_{T}}{\mSigma^{-1/2}} \vu}  \times  \left[ \vu^{H} \kron{\I_{T}}{\Hk} \vu \right] \nonumber \\
  &+ 2 \Q^{3/2} \Real{\dmudkH \kron{\I_{T}}{\mSigma^{-1/2}} \vu}  \times  \left[ \vu^{H} \kron{\I_{T}}{\Hj} \vu \right] \nonumber \\
  &+ \Q^{2} \left[ \vu^{H} \kron{\I_{T}}{\Hj} \vu \right] \left[ \vu^{H} \kron{\I_{T}}{\Hk} \vu \right].
\end{align}
Proceeding along the same lines as for the calculation of $T_{1}$, $T_{2}$ and $T_{3}$, it is straightforward to show that
\begin{align}
\tilde{T}_{1} &= \E{\Q \phi^{2}(\Q) \Real{\dmudjH \kron{\I_{T}}{\mSigma^{-1/2}} \vu}  \Real{\dmudkH \kron{\I_{T}}{\mSigma^{-1/2}} \vu}} \nonumber \\
&= \frac{1}{2MT} \E{\Q \phi^{2}(\Q)} \Real{\dmudjH \kron{\I_{T}}{\iSigma} \dmudk}
\end{align}
\begin{equation}
\tilde{T}_{2} = \E{\Q^{3/2} \phi^{2}(\Q) \Real{\dmudjH \kron{\I_{T}}{\mSigma^{-1/2}} \vu}  \times  \left[ \vu^{H} \kron{\I_{T}}{\Hk} \vu \right]} = 0
\end{equation}
and
\begin{align}
\tilde{T}_{3} &=\E{\Q^{2} \phi^{2}(\Q) \left[ \vu^{H} \kron{\I_{T}}{\Hj} \vu \right] \left[ \vu^{H} \kron{\I_{T}}{\Hk} \vu \right]} \nonumber \\
&= \frac{1}{M(MT+1)} \E{\Q^{2} \phi^{2}(\Q)} \left[ T \Tr{\Hj} \Tr{\Hk} + \Tr{\Hj \Hk} \right].
\end{align}
Gathering the previous equations, we finally obtain \textbf{the FIM for EVS distributions}:
\begin{align}\label{FIM_EVS}
\F(j,k) &= -T^{2} \Tr{\iSigma \Sigmaj} \Tr{\iSigma \Sigmak} \nonumber  \\
&+ \frac{2}{MT} \E{\Q \phi^{2}(\Q)} \sum_{t=1}^{T} \Real{\dmutdjH \iSigma \dmutdk} \nonumber \\
& + \frac{\E{\Q^{2} \phi^{2}(\Q)}}{M(MT+1)} \left[ T \Tr{\iSigma \Sigmaj} \Tr{\iSigma \Sigmak} + \Tr{\iSigma \Sigmaj \iSigma \Sigmak} \right].
\end{align}
Again, let us prove that when the data is Gaussian distributed, we recover the Slepian-Bangs formula. For Gaussian distributed data, one has
\begin{equation*}
p(\Q) = \frac{1}{\Gamma(MT)} \Q^{MT-1} \ex{-\Q}.
\end{equation*}
In this case, we have $\E{\Q \phi^{2}(\Q)} = MT$ and $\E{\Q^{2} \phi^{2}(\Q)} = MT(MT+1)$ and  \eqref{FIM_EVS} reduces to
\begin{align}
\Fg(j,k) &= -T^{2} \Tr{\iSigma \Sigmaj} \Tr{\iSigma \Sigmak} \nonumber  \\
&+ 2 \sum_{t=1}^{T} \Real{\dmudjH \iSigma \dmudk} \nonumber \\
& + T \left[T \Tr{\iSigma \Sigmaj} \Tr{\iSigma \Sigmak} + \Tr{\iSigma \Sigmaj \iSigma \Sigmak} \right]
\end{align}
which coincides with \eqref{Slepian-Bangs}.

\section{Conclusion}
In this letter, we proceeded to the extension of the well-known Slepian-Bangs formula of the FIM for Gaussian distributed data to the larger family of elliptically contoured distributions. Surprisingly enough, the new expression is rather simple and involves only slight modifications compared to its Gaussian counterpart. Only the expectation of some functions of the (scalar) modular variate are to be derived, which  can be done often analytically otherwise numerically. This result paves the way to derivation of new Cram\'{e}r-Rao bounds, e.g., for structured covariance matrix estimation or clutter and/or external noise parameters estimation in non-Gaussian environments.

\end{document}